\documentclass[prb,showpacs,amsmath,amssymb]{revtex4}

\usepackage{amsmath,amsthm,amssymb,amstext}
\usepackage{graphicx}
\usepackage{citesort}
\usepackage{enumerate}
\usepackage{array}

\newcommand{\dblqsum}{\tau}
\newcommand{\clim}{\lim_{*}}
\newcommand{\prefac}{h_{N,m,q}}
\newcommand{\colfn}{{\bf c}}
\newcommand{\afn}{ u}
\newcommand{\re}{\text{Re}}
\newcommand{\im}{\text{Im}}
\newcommand{\sad}{z_{+,i}}
\newcommand{\sadstar}{z_{+,i}^*}
\newcommand{\I}{I_{N,m,q}(\mu)}
\newcommand{\h}{\mathcal{H}_{N,m,q}(\mu)}
\newcommand{\nbhd}{\Omega_{+,i}}
\newcommand{\nbhdstar}{\Omega_{-,i}}
\newcommand{\nbhdboth}{\Omega_{\pm,i}}
\newcommand{\mult}{\cdot}

\newcommand{\supp}{\text{supp}}
\newcommand{\Arcsin}{\text{Arcsin}}
\newcommand{\Int}{\text{int}}

\newtheorem*{conj}{Conjecture}
\newtheorem{prop}{Proposition}
\newtheorem{lemma}{Lemma}

\begin{document}

\title{On the asymptotics of some large Hankel determinants generated by Fisher-Hartwig symbols defined on the real line}

\author{T. M. Garoni}
\affiliation{Institute for Mathematics and its Applications, 
University of Minnesota, 400 Lind Hall, 207 Church Street S.E., Minneapolis, MN 55455-0436}
\email{garoni@ima.umn.edu}
\homepage{http://www.ima.umn.edu/~garoni}

\date{\today}
\pacs{02.30.Mv,02.30.Gp,02.90.+p,02.50.Cw}

\begin{abstract}
We investigate the asymptotics of Hankel determinants of the form
$$
\det_{j,k=0}^{N-1}\left[\int_{\Omega}\,dx\,\omega_N(x)\,\prod_{i=1}^m\vert \mu_i-x\vert^{2q_i}\,x^{j+k}\right]
$$
as $N\to\infty$ with $q$ and $\mu$ fixed, where $\Omega$ is an infinite
subinterval of $\mathbb{R}$ and $\omega_N(x)$ is a positive
weight on $\Omega$. Such objects are natural analogues 
of Toeplitz determinants generated by Fisher-Hartwig symbols,
and arise in random matrix theory in the investigation of certain
expectations involving random characteristic polynomials.
The reduced density matrices of certain one-dimensional systems of trapped
impenetrable bosons can also be expressed in terms of Hankel
determinants of this form. 

We focus on the specific cases of scaled Hermite and Laguerre weights. 
We compute the asymptotics by using a duality formula expressing the $N\times N$ Hankel determinant
as a $2(q_1+\dots+q_m)$-fold integral, which is valid when each $q_i$ is natural.
We thus verify, for such $q$, a recent conjecture of Forrester and Frankel derived using a log-gas
argument. 
\end{abstract}

\maketitle

\section{Introduction}
Consider the multiple integral
\begin{equation}
H_{M,N,m,q}(\mu):=\int_{\Omega}\omega_N(z_1)dz_1\dots\int_{\Omega}\omega_N(z_M)dz_M
\, \vert\Delta_M(z)\vert^2\, \prod_{l=1}^M\prod_{i=1}^m\vert \mu_i-z_l\vert^{2q_i},
\label{multiple integral definition}
\end{equation}
where
\begin{equation}
q:=(q_1,\dots,q_m),
\qquad \qquad 
\mu:=(\mu_1,\dots,\mu_m),
\end{equation}
$\omega_N(z)$ is a nonzero and
continuous weight function, possibly depending on a parameter $N$, and 
\begin{equation}
\Delta_M(z):=\det_{j, k=1}^M(z_k^{j-1})=\prod_{1\le j<k\le M}(z_k-z_j)
\label{Vandermonde}
\end{equation}
is the Vandermonde determinant. As a notational convenience we also define
\begin{align}
H_{M,N}&:=H_{M,N,0,\cdot}(\cdot),\\
&\phantom{:}=\int_{\Omega}\omega_N(z_1)dz_1\dots\int_{\Omega}\omega_N(z_M)dz_M
\, \vert\Delta_M(z)\vert^2.
\end{align}

It is known that $M$-fold integrals of the form 
(\ref{multiple integral definition}) can be identified with the
determinant of an $M\times M$ matrix. Expanding the
Vandermonde determinants in terms of sums over permutations, and
simplifying appropriately, we find
\begin{equation}
H_{M,N,m,q}(\mu)=M!\,\det_{j,k=0}^{M-1}\left[\int_{\Omega}dz\,a(z)\,z^j\,(z^*)^k\right],
\label{Heine identity}
\end{equation}
where $z^*$ denotes the complex conjugate of $z$, and 
\begin{equation}
a(z):=\omega_N(z)\,\prod_{i=1}^m\vert \mu_i-z\vert^{2q_i}\label{Fisher-Hartwig symbol}.
\end{equation}
When
$\Omega\subseteq\mathbb{R}$, equation (\ref{Heine identity}) becomes
\begin{equation}
H_{M,N,m,q}(\mu)
=M! \,
\det_{j,k=0}^{M-1}
\left[
a_{j+k}
\right],
\label{Hankel}
\end{equation}
where
\begin{equation}
a_n:=\int_{\Omega}dz\,a(z) \,\,z^{n}.
\end{equation}
One says that the determinant (\ref{Hankel}) is {\em
  generated} by the function (\ref{Fisher-Hartwig symbol}).

Since the entries in the determinant (\ref{Hankel}) are of the form $a_{j+k}$, we have thus 
identified the multiple integral (\ref{multiple integral definition}), when $\Omega\subseteq\mathbb{R}$, 
with a Hankel determinant. Had we instead taken $\Omega$ to be $\mathbb{T}$, the
unit circle in $\mathbb{C}$, then according to (\ref{Heine identity}) the entries of the determinant would
be $a_{j-k}$, and we would thus obtain a Toeplitz
determinant. (It is conventional when discussing such Toeplitz
determinants to set $z=e^{i\theta}$ and to define $a_n$ as the
integral of $a(z)\,z^n$ with respect
to $d\theta$ rather than $dz$. This merely introduces the nonzero factor
$i\,e^{i\theta}$ which is easily absorbed into the definition of
$a(z)$, and so such technicalities are not relevant to our discussion here.)

When $z\in\mathbb{T}$, functions of the form (\ref{Fisher-Hartwig symbol}) are known as
{\em Fisher-Hartwig symbols} \cite{BottcherSilbermann}(although we remark that they are not the most
general examples of Fisher-Hartwig symbols). By extension, we can describe the Hankel determinant
(\ref{Hankel}) as being generated by a Fisher-Hartwig symbol which is defined on the
real line. 

The asymptotic analysis of Toeplitz
determinants generated by Fisher-Hartwig symbols is a fascinating and
well studied subject (see e.g. \cite{BottcherSilbermann} and
references therein), and rigorous results which describe the
large $M$ asymptotic behavior of Toeplitz determinants generated by symbols of
the form (\ref{Fisher-Hartwig symbol}) are known
\cite{Widom73}. There are a number of important physical
applications of such determinants (see e.g. \cite{FisherHartwig68,Lenard72,ForresterFrankel04}).
It is often the case, as we discuss presently, that the 
quantity appearing in applications is
actually the integral (\ref{multiple integral definition}) rather than
the determinant directly, and 
 when 
$\Omega\subseteq\mathbb{R}$ we are naturally lead to Hankel
determinants generated by the 
symbol (\ref{Fisher-Hartwig symbol}). We discuss below a number of
physical applications in which the asymptotics of such Hankel determinants is
of interest.
A rigorous treatment of these asymptotics is an open problem, 
however a conjectured form for the large $M$
asymptotics has recently been
reported by Forrester and Frankel in \cite{ForresterFrankel04}. 
Very recently, the $m=1$ case of this conjecture
has been verified when $\Omega=\mathbb{R}$ and $\omega_N(x)$ is a
Hermite weight, by using a Riemann Hilbert approach \cite{Krasovsky04}.
In the present work we verify the conjecture of Forrester and Frankel 
when $\omega_N(z)$ is either a Hermite or Laguerre weight for any
$m\in\mathbb{N}$, when each $q_i\in\mathbb{N}$. 

\subsection{Random matrix theory}
The multiple integral (\ref{multiple integral definition}) has a natural 
interpretation in random matrix theory.
Let us consider the ensemble of random matrices with joint eigenvalue probability density function (pdf) given by
\begin{equation}
P_{M,N}(x):=\frac{1}{H_{M,N}}\,\,|\Delta_M(x)|^2\,\prod_{l=1}^M \omega_N(x_l),
\label{eigenvalue density}
\end{equation}
and whose eigenvalues lie in $\Omega$.
When $\Omega=\mathbb{R}$, concrete examples of such ensembles include 
the ubiquitous Gaussian unitary ensemble (GUE), corresponding to ${\omega_N(x)=\exp(-2N x^2)}$, 
as well as more general unitary ensembles (UE), corresponding to ${\omega_N(x)=\exp(-N V(x))}$ with $V(x)$ an arbitrary 
polynomial of even degree with positive leading coefficient
(see e.g. \cite{Deift}). 
When ${\Omega=(0,\infty)}$ an important 
example is the Laguerre unitary ensemble (LUE), corresponding to ${\omega_N(x)=x^{\alpha}\exp(-4N x)}$, 
which includes Wishart matrices and the Chiral GUE as special cases (the latter after a straightforward
change of variables; see e.g. \cite{ForresterBook}). Setting 
$x=e^{i\theta}$ and $\omega_N(x)=1$ in
(\ref{eigenvalue density}) we obtain the joint pdf for the eigenphases 
$\theta_l\in [0,2\pi]$    
of the ensemble of random
unitary matrices with Haar measure, often called the circular unitary
ensemble (CUE). For the purpose of computing expectations, the CUE is
equivalent to (\ref{eigenvalue density}) with $\Omega=\mathbb{T}$ and $\omega_N(z)=1/i\,z$.

If we denote the characteristic polynomial of the $M\times M$ matrix $X$, with eigenvalues $x_1,\dots,x_M$, by
\begin{equation}
\mathcal{Z}_M(\mu_i):=\det(\mu_i\,I-X)=\prod_{l=1}^M(\mu_i-x_l),
\label{characteristic polynomial}
\end{equation}
then (\ref{multiple integral definition}) corresponds to the following
expectation involving
the absolute value of 
such characteristic 
polynomials
\begin{equation}
\left\langle\prod_{i=1}^m \vert\mathcal{Z}_M(\mu_i)\vert^{2q_i} \right\rangle_{P_{M,N}} = 
\frac{H_{M,N,m,q}(\mu)}{H_{M,N}},
\label{H ratio as characteristic polynomial correlation}
\end{equation}
where the expectation on the left hand side of (\ref{H ratio as characteristic polynomial correlation}) is with respect 
to the joint eigenvalue pdf (\ref{eigenvalue density}). 
The case $M=N$ is generally the case of interest.

From
(\ref{Heine identity}) we see that expectations of characteristic
polynomials of the form appearing in (\ref{H ratio as characteristic polynomial correlation}) 
are characterized by a determinant generated by the symbol
(\ref{Fisher-Hartwig symbol}); when $\Omega\subseteq\mathbb{R}$, 
it is a Hankel determinant, and when $\Omega=\mathbb{T}$ it is a Toeplitz
determinant. 

A sizable literature on the correlations of products and ratios of characteristic polynomials of random 
matrices from various
ensembles has emerged in recent years, see e.g.
\cite{BrezinHikami00,KeatingSnaith00,MehtaNormand01,StrahovFyodorov03,FyodorovKeating03,AkemannFyodorov03,ForresterKeating04,BaikDeiftStrahov03}, 
and significant progress has been made in the understanding of such
objects. Such quantities have applications in diverse fields including
number theory, quantum chaos and many-body quantum mechanics.
These works consider either exact algebraic relations that are valid for finite $N$, or the large $N$ 
asymptotics in the usual universal microscopic scaling limits.
We shall be interested not in scaling limits, but in the limit of
large $N$ with $\mu$ fixed.
Investigations of objects of the form
(\ref{H ratio as characteristic polynomial correlation}) in this limit have been reported in
\cite{ForresterFrankel04,ForresterFrankelGaroni03}.

\subsection{Impenetrable bosons}
A compelling physical motivation for investigating multiple integrals of the form 
(\ref{multiple integral definition}) arises from a consideration of
certain one-dimensional many-body systems of impenetrable bosons.
By impenetrability we simply mean that we require the wavefunction to
vanish whenever two bosons occupy the same point in space. Such systems have been receiving renewed theoretical interest 
recently due to the possibility of their experimental realization in the near future
using ultra-cold systems of atomic bosons confined in
elongated traps; see e.g. \cite{Olshanii98,DunjkoLorentOlshanii01}. 
Systems of impenetrable bosons with certain specific boundary
conditions are known to have ground state wavefunctions 
of the form
\begin{equation}
\psi(x_1,\dots,x_M)= \frac{1}{\sqrt{C_M}}\prod_{l=1}^M\sqrt{g_N(x_l)}\,\, \vert\Delta_M[f(x_1),\dots,f(x_M)]\vert,
\qquad x_1,\dots,x_M\in D\subseteq \mathbb{R},
\label{wavefunction}
\end{equation}
see e.g. \cite{ForresterFrankelGaroniWitte03a,TimPhD}. Specifically, systems with periodic,
Dirichlet, or Neumann boundary conditions have
wavefunctions of this form, as
do systems confined in an harmonic well.
Of these, the harmonically confined system is perhaps the most
relevant to current experiments.

It is worth emphasizing that the introduction of 
such zero-range infinite-strength interactions establishes a correspondence 
between impenetrable bosons and a corresponding system of free fermions, 
a fact first noted in \cite{Girardeau60}, and this is one
of the primary reasons for current experimental interest in such
systems. Indeed, for the specific systems mentioned above,
were it not for the absolute value surrounding the Vandermonde determinant, (\ref{wavefunction}) 
would define the wavefunction for a system of free fermions. This fact implies that certain quantities such as the 
energy spectrum and the particle density are identical in the impenetrable boson system 
and its corresponding free fermion system. 
Quantities which depend on the phase of the wavefunction however will clearly differ significantly between these two
systems.

One such quantity, of great significance,
is the $n$-body density matrix, which for a system of $N+n$ particles
is defined as
\begin{equation}
\rho_{N+n}^{(n)}(x_1,y_1,\dots,x_n,y_n)
:=\binom{N+n}{n}
\int_{D}d\xi_{1}\dots\int_{D}d\xi_{N}\,
\psi(x_1,\dots,x_n,\xi_{1},\dots,\xi_{N})
\psi^*(y_1,\dots,y_n,\xi_{1},\dots,\xi_{N}).
\label{density matrix definition}
\end{equation}
A key observation is that wavefunctions of the form (\ref{wavefunction}) admit the factorization
\begin{equation}
\psi(x_1,\dots,x_n,\xi_1,\dots,\xi_{N})
=\sqrt{\frac{C_{N}}{C_{N+n}}}\,\,\prod_{l=1}^n\sqrt{g_N(x_l)}\,\,\vert\Delta_n[f(x_1),\dots,f(x_n)]\vert
\prod_{l=1}^N\prod_{i=1}^{n}\vert f(x_i)-f(\xi_l)\vert\,\mult\,
\psi(\xi_1,\dots,\xi_{N}),
\label{wavefunction factorisation}
\end{equation}
and inserting (\ref{wavefunction factorisation}) into 
(\ref{density matrix definition}) yields
\begin{equation}
\begin{split}
\rho_{N+n}^{(n)}(x_1,y_1,\dots,x_n,y_n)
&
=\binom{N+n}{N}\,\prod_{i=1}^n\sqrt{g_N(x_i)\,g_N(y_i)}
\vert\Delta_n[f(x_1),\dots,f(x_n)]\Delta_n[f(y_1),\dots,f(y_n)]\vert
\\&\quad\times
\frac{1}{H_{N+n,N}}H_{N,N,2n,q}(f(x_1),\dots,f(x_n),f(y_1),\dots,f(y_n))\Big\vert_{q=(1/2,\dots,1/2)},
\label{density matrix in terms of H}
\end{split}
\end{equation}
where in the definition of $H_{N,N,2n,q}$  and $H_{N+n,N}$ we have
$\Omega=f(D)$ and 
\begin{equation}\omega_N(z)= g_N(f^{-1}(z))\,\frac{df^{-1}}{dz}(z).
\label{effective impenetrable boson weight}
\end{equation}
For the four specific systems mentioned below (\ref{wavefunction}), 
this $\omega_N(z)$ given in (\ref{effective impenetrable boson weight}) is
well defined, nonzero and continuous.

Hence, from (\ref{Heine identity}) we see that $\rho_{N+n}^{(n)}$
is characterized by
a determinant generated by the symbol (\ref{Fisher-Hartwig symbol}). 
For systems subject to periodic boundary conditions 
this determinant will be a Toeplitz determinant \cite{Lenard64,ForresterFrankelGaroniWitte03b}, whereas for systems
confined by an harmonic well it will be a Hankel
determinant \cite{ForresterFrankelGaroniWitte03b}. Indeed, this link between density matrices for
impenetrable bosons and Toeplitz determinants generated by
Fisher-Hartwig symbols was originally one of the key motivations for
investigating the asymptotics of such Toeplitz determinants
\cite{Lenard64,Lenard72,FisherHartwig68}. 
In light of the possible future experimental realization of finite one-dimensional
harmonically trapped systems of impenetrable bosons, 
an important theoretical question is the behavior of the corresponding  
density matrices when $N$ is large and $x_1,y_1,\dots,x_n,y_n$ are fixed.
This then provides a direct physical motivation for investigating the
large $N$ asymptotics of $N\times N$ Hankel determinants generated by
symbols of the form (\ref{Fisher-Hartwig symbol}).

Perhaps the most important quantity is the one-body density matrix. 
The asymptotics of the one-body density matrix 
for a system with periodic boundary conditions can be rigorously established
from the asymptotics of the corresponding Toeplitz determinant. 
The leading order behavior of the one-body density matrix in the case of harmonic confinement was 
deduced in \cite{ForresterFrankelGaroniWitte03b} using log-gas arguments, and has been recovered in 
\cite{Gangardt04} using a rather more direct, yet still non-rigorous, approach.
The asymptotics of the one-body density matrix in the Dirichlet/Neumann case was deduced in 
\cite{ForresterFrankelGaroni03}, again by log-gas arguments, and has now
been rigorously proved in \cite{ForresterFrankel04}, by making use of recent results in 
\cite{BasorEhrhardt02,BasorEhrhardt01}. 

We conclude our discussion of impenetrable bosons by noting the correspondence
between the joint eigenvalue pdf (\ref{eigenvalue density}) 
and the wavefunction (\ref{wavefunction}).
The correspondence between the joint eigenvalue pdf 
(\ref{eigenvalue density}) and the wavefunction for a system of free fermions is well
known \cite{Mehta}.
The correspondence between impenetrable bosons and random matrices was first noted by Sutherland \cite{SutherlandPRA71}, 
between systems of impenetrable bosons with periodic 
boundary conditions and the CUE; see also \cite{SutherlandJMP71}.
The correspondence between impenetrable bosons with Dirichlet or Neumann boundary conditions and the 
Jacobi unitary ensembles (JUE) was discussed in
\cite{ForresterFrankelGaroniWitte03a,ForresterFrankelGaroni03}, and a similar interpretation for the LUE was noted in 
\cite{ForresterFrankel04}. Again, the most interesting case from an experimental perspective 
is the correspondence between the GUE and systems of impenetrable bosons confined in an harmonic well, and 
this particular system has been the focus of considerable recent theoretical study, see e.g.
\cite{ForresterFrankelGaroniWitte03b,Papenbrock,Gangardt04} and references therein.

\subsection{Asymptotics of Hankel determinants}
The asymptotics of large Toeplitz and Hankel determinants
has been of long standing interest to mathematicians. 
For Toeplitz determinants generated by well behaved symbols, very precise
asymptotic results are given by the Szeg\"o limit theorems
(see e.g. \cite{BottcherSilbermann,Johansson88}). 
Toeplitz determinants generated by the symbol (\ref{Fisher-Hartwig symbol}) are not amenable to the Szeg\"o limit
theorems however since (\ref{Fisher-Hartwig symbol}) has zeros.
Inspired in part by
applications to impenetrable bosons Lenard \cite{Lenard72,Lenard64}
(see also \cite{FisherHartwig68}) conjectured the asymptotics of Toeplitz determinants
generated by symbols of the form (\ref{Fisher-Hartwig symbol}),
and this conjecture was subsequently proved by Widom
\cite{Widom73}. 
The asymptotic behavior of Toeplitz determinants generated by
(\ref{Fisher-Hartwig symbol}), as well as more general
Fisher-Hartwig symbols, is  now 
well understood (see e.g. \cite{BottcherSilbermann}). 
Analogously, the asymptotic behavior of large Hankel determinants 
generated by functions defined on $\Omega\subseteq\mathbb{R}$ has also been the subject of study.
This problem was addressed by Szeg\"o 
\cite{SzegoHankelForms} and also Hirschman \cite{Hirschman66} with $\Omega$ a finite interval 
(see also \cite{Johansson88}). In the context of the UE and LUE of random matrix theory, as well as in the context 
of trapped systems of impenetrable bosons, we are interested in case
where $\Omega$ is infinite, and 
recently Basor {\em et al} \cite{BasorChenWidom01} have considered the
asymptotics of Hankel determinants generated by symbols defined on $\Omega=(0,\infty)$.
However, a key restriction in these works is that the symbol be nowhere
zero, and hence they do not apply to determinants generated by
(\ref{Fisher-Hartwig symbol}). 
Forrester and Frankel\cite{ForresterFrankel04}
have recently conjectured the asymptotic behavior of 
Hankel determinants generated by symbols of the form 
(\ref{Fisher-Hartwig symbol}) defined on
$\Omega\subseteq\mathbb{R}$. Complete and rigorous
proofs of their conjectures remains an open problem. As mentioned above, a rigorous proof for the $m=1$
case when $\Omega=\mathbb{R}$ and $\omega_N(x)$ is a Hermite
weight has very recently been reported in \cite{Krasovsky04}.

Specifically, Forrester and Frankel \cite{ForresterFrankel04} consider
the behavior of  
the ratio
\begin{equation}
\h:=
\frac{H_{N,N,m,q}(\mu)}
{H_{N+|q|,N}},
\label{ratio}
\end{equation}
where
\begin{equation}
|q|:=q_1+\dots +q_m,
\label{partition definition}
\end{equation}
as $N\to\infty$ with $\mu$
and $q$ fixed.
Amongst other results, they consider the case ${\omega_N(x)=e^{-N \,V(x)}}$
with either ${\Omega=\mathbb{R}}$ or ${\Omega=(0,\infty)}$. For a
particular choice of such $\omega_N(x)$ 
and $\Omega$
there corresponds the quantity $\rho(x)$, which, with $P_{N,N}$
defined as in (\ref{eigenvalue density}), equals the limit of
\begin{equation}
\int_{\Omega^{N-1}}P_{N,N}(x,x_2,\dots,x_N)\,dx_2\dots dx_N
\end{equation}
as $N\to \infty$ with $x$ held fixed;
i.e. $\rho(x)$ is the limiting expected
eigenvalue density of the ensemble of random matrices defined by $P_{N,N}$. 
We note that $\rho(x)$ is non-negative and has compact support.
In what follows $\Int(\supp\,\rho)$ denotes the interior of the
support of $\rho(x)$.
For a
detailed discussion of $\rho(x)$, and a number of other very interesting alternative
characterizations of $\rho(x)$, the reader is referred to \cite{Deift}.
The conjecture reported in \cite{ForresterFrankel04} (in our notation) is the following.
\begin{conj}[Forrester-Frankel]
Let $ m,N \in \mathbb{N}$, 
$q\in (-1/2,\infty)^m$ 
and 
$\mu_1,\dots,\mu_m \in {\rm int} ({\rm supp}\,\,\rho)$, where $\rho(x)$ 
is defined as above.
Furthermore, suppose either ${\Omega=\mathbb{R}}$ or ${\Omega=(0,\infty)}$, and take ${\omega_N(x)=e^{-N \,V(x)}}$
where $V(x)$ is a polynomial which is independent of $N$ and which has positive leading coefficient 
and no zeros in $\Omega$. Then 
\begin{equation}
\begin{split}
\h
&
=N^{\sum_{i=1}^m(q_i^2-q_i)}
\prod_{i=1}^m[\omega_N(\mu_i)]^{-q_i}
\prod_{i=1}^m 
\frac{G^2(q_i+1)}{G(2q_i+1)}(2\pi)^{q_i^2-q_i}
\\&\quad\times
\prod_{1\le j<k\le m}\vert \mu_k-\mu_j\vert^{-2q_j q_k}\prod_{i=1}^m[\rho(\mu_i)]^{q_i^2}
\,\left[1+o(1)\right].
\end{split}
\label{FF conjecture equation}
\end{equation}
\label{FF conjecture}
\end{conj}
\noindent Here and in the sequel $G$ refers to Barnes' $G$-function \cite{Barnes00}.

Despite first appearances, the structure of (\ref{FF conjecture equation}) 
is actually quite simple. Note that the only ensemble dependent quantities on the right
hand side are $\rho(x)$ and $\omega_N(x)$, 
and that the dependence of $\mathcal{H}_{M,m,q}(\mu)$ on these two quantities is 
universal.  
The other quantities on the right hand side are truly universal. 
We note that the factor $G^2(q_i+1)/G(2q_i+1)$ occurs
also in the asymptotics of Toeplitz determinants generated by
Fisher-Hartwig symbols \cite{Widom73}, and has been discussed in the
context of moments of
random characteristic polynomials
\cite{KeatingSnaith00,BrezinHikami00}. 
We should note that the actual conjecture reported in \cite{ForresterFrankel04} is slightly 
more general than (\ref{FF conjecture equation}), but (\ref{FF conjecture equation}) is sufficient for our purposes.

As an aside, we remark that the ratio appearing in 
(\ref{density matrix in terms of H}) is precisely of the form
(\ref{ratio}) and so 
the asymptotic behavior of all the $n$-body density matrices
for a system of harmonically confined impenetrable bosons follows
directly from (\ref{FF conjecture equation}).

The present article focuses on two classical
cases already
mentioned, the case of the Hermite weight, corresponding to the GUE,
and the case of the Laguerre weight, corresponding to the LUE.
In the Hermite case
\begin{equation}
\omega_N(x)=e^{-2Nx^2},\qquad\quad
\Omega=\mathbb{R},\qquad\quad
\rho(\mu)=\frac{2}{\pi}\sqrt{1-\mu^2},\qquad\quad
\text{supp}(\rho)=[-1,1],
\label{Hermite weight}
\end{equation}
and in the Laguerre case
\begin{equation}
\omega_N(x)=x^{\alpha}\,e^{-4Nx},\qquad\quad
\Omega=(0,\infty),\qquad\quad
\rho(\mu)=\frac{2}{\pi}\sqrt{\frac{1}{\mu}-1},\qquad\quad
\text{supp}(\rho)=[0,1].
\label{Laguerre weight}
\end{equation}
The conjecture reported in \cite{ForresterFrankel04}
was deduced by considering the specific examples of the Hermite and Laguerre 
cases, to find the general form in terms of $\rho(x)$ and
$\omega_N(x)$. The asymptotics for these two cases
was deduced by using the log-gas analogy to conjecture a factorization of (\ref{ratio}), computing the 
asymptotics of each factor when the $q_i$ were natural, and then conjecturing an analytic continuation to
real $q_i$. 

In this work we show how (\ref{FF conjecture equation})
can be proved rigorously for the Hermite and Laguerre cases, 
when each $q_i$ is natural, by using a duality formula 
derived from a general result in \cite{BrezinHikami00}. 
By a {\em duality}  formula we mean an equation identifying the $N$-fold integral $\h$
with a {$2|q|$~-~fold} integral.
Our results clarify the origin of the factors 
appearing in (\ref{FF conjecture equation}).
A similar approach has been used in \cite{Gangardt04} 
to investigate a particular special case when the weight was of Hermite type, in the context of trapped impenetrable 
bosons.

Section \ref{Duality formula} contains a discussion of the duality formula. Due to  the similar nature of the 
Laguerre and Hermite cases, they can both be derived simultaneously. 
Section \ref{Asymptotics} then 
discusses the asymptotic analysis of the {$2|q|$~-~fold} integral obtained from the duality 
formula by means of the saddle point method. 
We show how to deduce the general form of all terms and explicitly simplify
the leading order term, and thus verify the conjecture 
(\ref{FF conjecture equation}) for the Hermite and Laguerre cases when each $q_i$ is natural.

\section{Duality formula}
\label{Duality formula}
Let us define
\begin{equation}
F_{K,N}(\lambda_1,\dots,\lambda_K):=\frac{1}{\Delta_K(\lambda_1,\dots,\lambda_K)}
\det_{j,k=1}^K\pi^{(N)}_{N+j-1}(\lambda_k),
\label{Brezin Hikami F}
\end{equation}
where $\{\pi_k^{(N)}\}_{k=0}^{\infty}$ are the monic orthogonal polynomials corresponding to
$\omega_N(x)$ and $\Omega$; i.e. they are uniquely defined by the following two conditions:
\begin{equation}
\int_{\Omega} \,\omega_N(x) \, dx\,\,\pi^{(N)}_j(x) \pi^{(N)}_k(x)=0,\qquad j\neq k
\end{equation}
\begin{equation}
\pi^{(N)}_j(x)=x^j+O(x^{j-1}).
\end{equation}
For the Hermite and Laguerre cases, the $\pi_{N+j-1}^{(N)}(x)$ can be expressed in terms of the standard Hermite and
Laguerre polynomials found in  Szeg\"o's classic book\cite{Szego} as follows
\begin{equation}
\pi^{(N)}_{N+j-1}(x)=
\begin{cases}
2^{-3(N+j-1)/2}N^{-(N+j-1)/2}H_{N+j-1}(\sqrt{2N}x), & \text{Hermite,}
\\
(-1)^{N+j-1}(N+j-1)!(4N)^{-N-j+1}\, L_{N+j-1}^{(\alpha)}(4Nx), & \text{Laguerre.}
\\
\end{cases}
\end{equation}

According to Br\'ezin and Hikami \cite{BrezinHikami00}, we have the
following very useful identity
\begin{equation}
F_{K,N}(\lambda_1,\dots,\lambda_K)
=
\frac{1}{H_{N,N}} \, \int_{\Omega}\omega_N(x_1)dx_1\dots\int_{\Omega}\omega_N(x_N)dx_N\,\,\Delta^2_N(x) 
\,\,\prod_{l=1}^N\prod_{i=1}^K
(\lambda_i-x_l).
\label{brezin-hikami identity}
\end{equation}
If we restrict ourselves to $q\in\mathbb{N}^m$ and set 
\begin{equation}
K=2q_1+\dots+2q_m=2|q|,
\label{K definition}
\end{equation}
we can consider the confluent limit 
\begin{equation}
\clim:=
\lim_{\lambda_{\dblqsum(m)+2q_m}\to \mu_m}
\dots
\lim_{\lambda_{\dblqsum(m)+1}\to \mu_m}
\dots
\lim_{\lambda_{2q_1+2q_2}\to \mu_2}
 \dots
\lim_{\lambda_{2q_1+1}\to \mu_2}\,\,
\lim_{\lambda_{2q_1}\to \mu_1}
\dots
\lim_{\lambda_1\to \mu_1},
\label{confluence}
\end{equation}
where 
\begin{equation}
\dblqsum(i):=\sum_{l=1}^{i-1}2q_l.
\end{equation}
Taking the limit (\ref{confluence}) of both sides of (\ref{brezin-hikami identity}), and using the elementary fact that 
$(\mu_i-x_l)^{2q_i}=\vert\mu_i-x_l\vert^{2q_i}$ when $q_i$ is an integer and $\mu_i$ and $x_l$ are real,
we thus obtain
\begin{equation}
\clim\, F_{K,N}(\lambda_1,\dots,\lambda_K)=\frac{H_{N,N,m,q}(\mu)}{H_{N,N}},
\end{equation}
and therefore
\begin{equation}
\h=\frac{H_{N,N}}{H_{N+|q|,N}}\,\,\clim\, F_{K,N}(\lambda_1,\dots,\lambda_K), 
\qquad\qquad q\in\mathbb{N}^m.
\end{equation}
This is the key relation we need to derive the duality formula for $\h$, all that remains is to 
take the confluent limit of $F_{K,N}$. 

For later convenience, we set
\begin{equation}
\pi^{(N)}_{N+j-1}(x)=\zeta_N(x)\, r_{N+j-1}(x).
\label{pi as r}
\end{equation}
If we insert (\ref{pi as r}) into (\ref{Brezin Hikami F}) and take the limit (\ref{confluence}), then by factoring 
the Vandermonde determinant we obtain
\begin{align}
\clim F_{K,N}(\lambda_1,\dots,\lambda_K)
&=\clim \prod_{1\le j<k\le m}\,\,\prod_{l_k=1}^{2q_k}\prod_{l_j=1}^{2q_j}
(\lambda_{\dblqsum(k)+l_k}-\lambda_{\dblqsum(j)+l_j})^{-1}
\,\,
\prod_{j=1}^K\zeta_N(\lambda_j)
\nonumber
\\&\quad\times
\clim
\prod_{i=1}^m\Delta_{2q_i}^{-1}(\lambda_{\dblqsum(i)+1},\dots,\lambda_{\dblqsum(i+1)})
\det_{j,k=1}^K r_{N+j-1}(\lambda_k),
\\
&=
\prod_{1\le j<k\le m}(\mu_k-\mu_j)^{-4q_k q_j}\,\,\prod_{i=1}^m\zeta_N^{2q_i}(\mu_i)
\nonumber
\\
&\quad\times
\clim
\prod_{i=1}^m\Delta_{2q_i}^{-1}(\lambda_{\dblqsum(i)+1},\dots,\lambda_{\dblqsum(i+1)})
\det_{j,k=1}^K r_{N+j-1}(\lambda_k).
\label{pre Vandermonde L'Hopital confluence F limit}
\end{align}
To compute the remaining limit in (\ref{pre Vandermonde L'Hopital confluence F limit}) we can use
the following.
\begin{lemma}
\label{Vandermonde L'Hopital confluence}
Let $\colfn(\lambda)$ denote a column vector with $\dblqsum(m+1)$
entries, then for $i=1,2,\dots,m$ 
\begin{multline}
\lim_{\lambda_{\dblqsum(i)+2q_i}\to \mu_i} \dots \lim_{\lambda_{\dblqsum(i)+1}\to \mu_i}
\frac{1}{\Delta_{2q_i}(\lambda_{\dblqsum(i)+1},\dots,\lambda_{\dblqsum(i)+2q_i})}\,
\\\times
G(2q_i+1)
\det\left[ \colfn(\lambda_1)\dots \colfn(\lambda_{\dblqsum(i)+1}) \dots \colfn(\lambda_{\dblqsum(i)+2q_i})
\dots \colfn(\lambda_{\dblqsum(m+1)})
\right]
\\=
\det\left[ 
\colfn(\lambda_1)
\dots 
\colfn(\lambda_{\dblqsum(i)})
\,\,
\colfn(\mu_{i})
\,
\frac{d}{d \mu_i}\colfn(\mu_i)
\dots 
\frac{d^{2q_i-1}}{d \mu_i^{2q_i-1}}\colfn(\mu_i)
\,\,
\colfn(\lambda_{\dblqsum(i+1)+1})
\dots \colfn(\lambda_{\dblqsum(m+1)})
\right],
\end{multline}
where $G$ is Barnes' $G$-function.
\end{lemma}
\begin{proof}
This is easily proven by induction using L'H\^opital's rule, and recalling the identity
\begin{equation}
\prod_{l=1}^n\Gamma(l)=G(n+2).
\end{equation}
\end{proof}
Applying Lemma \ref{Vandermonde L'Hopital confluence} to (\ref{pre Vandermonde L'Hopital confluence F limit}) 
independently for each set
${\{\lambda_{\dblqsum(i)+1},\dots,\lambda_{\dblqsum(i)+2q_i}\}}$ with ${i=1,2,\dots,m}$,  results in
\begin{equation}
\clim F_{K,N}(\lambda_1,\dots,\lambda_{K})=
\prod_{1\le j<k\le m}(\mu_k-\mu_j)^{-4q_k q_j}\,\,\prod_{i=1}^m\frac{\zeta_N^{2q_i}(\mu_i)}{G(2q_i+1)}
\det_{\substack{1\le l_i \le 2q_i \\ 1\le i\le m\\1\le j\le 2|q|}}
\left[\frac{d^{l_i-1}}{d\mu_i^{l_i-1}}r_{N+j-1}(\mu_i)\right].
\label{q-det form of lim F}
\end{equation}
In (\ref{q-det form of lim F}) the columns of the determinant are
ordered such that one starts with $i=1$, writes out the $2q_1$ columns
depending on $l_1$, and then moves to $i=2$ etc.
We remark that we have now 
already obtained one of the two Barnes $G$-function factors that
appear in (\ref{FF conjecture equation}). 

The special property possessed by the Hermite and Laguerre polynomials that allows us to derive a 
duality formula for $\h$ for the specific weights 
(\ref{Hermite weight}) and (\ref{Laguerre weight}) is that they can be
expressed in 
terms 
of contour integrals. Indeed, by suitably massaging the 
standard results in Szeg\"o's book \cite{Szego} we find
\begin{align}
\pi_{N+j-1}^{(N)}(\mu_i)
&=
\begin{cases}
\begin{displaystyle}
c_j(N)
\,
e^{2N\,\mu_i^2}
\,
\int_{\mathcal{C}}dz\,
e^{-2N z\, \mu_i\,+\,Nz^2/2}
z^{N+j-1}
\end{displaystyle}
& 
\text{Hermite},\\
\begin{displaystyle}
c_j(N)
 \int_{\mathcal{C}}dz\,
e^{-2N z\, \mu_i}\frac{(z+2)^{N+\alpha}}{z^{N+1}}\left(\frac{1}{z}+\frac{1}{2}\right)^{j-1}
\end{displaystyle}
&
\text{Laguerre},\\
\end{cases}
\label{pi integrals}
\\
c_j(N)
&= 
\begin{cases}
\begin{displaystyle}
\sqrt{\frac{2N}{\pi}}\frac{1}{i\,2^{N+j}} 
\end{displaystyle}
& \qquad\qquad\text{Hermite},\\
\begin{displaystyle}
(-1)^{N+j-1}\frac{(N+j-1)!}{N^{N+j-1}}\frac{1}{2^{2N+j+\alpha}\pi i}
\end{displaystyle}
&
\qquad\qquad\text{Laguerre},\\
\end{cases}
\end{align}
where in the Hermite case the contour $\mathcal{C}$ lies along the imaginary axis and is oriented from 
$-i\infty$ to $+i\infty$, and
in the Laguerre case $\mathcal{C}$ is a closed positively oriented contour 
which encircles the origin but does not contain the point 
$z=-2$.

It is now straightforward to compute the derivatives required in (\ref{q-det form of lim F}) from the contour integrals
in (\ref{pi integrals}). Defining 
\begin{equation}
\zeta_N(\mu_i)=
\begin{cases}
e^{2N\, \mu_i^2} & \text{Hermite,}\\
1 & \text{Laguerre,}\\
\end{cases}
\end{equation}
and recalling the definition (\ref{pi as r}) we obtain
\begin{equation}
\frac{d^{l_i-1}}{d\mu_i^{l_i-1}}r_{N+j-1}(\mu_i)
=
c_j(N) \,d_{l_i}(N) \int_{\mathcal{C}}dz \,\,e^{-NS(z,\mu_i)}\,\afn(z)\,z^{l_i-1}[z^{\delta}+c]^{j-1},
\label{general form of the r derivatives}
\end{equation}
where
\begin{align}
\label{action}
S(z,\mu_i)
&:=
\begin{cases}
\begin{displaystyle}
2 \mu_i\, z - \log(z) -\frac{z^2}{2}
\end{displaystyle}
& 
\text{Hermite},\\
\begin{displaystyle}
2 \mu_i\, z + \log(z)-\log(z+2)
\end{displaystyle}
&
\text{Laguerre},\\
\end{cases}
\\
\afn(z)
&:=
\begin{cases}
1 & \qquad\qquad\qquad\qquad\text{Hermite,}\\
\begin{displaystyle}
\frac{(z+2)^{\alpha}}{z}
\end{displaystyle}
&
\qquad\qquad\qquad\qquad\text{Laguerre,}\\
\end{cases}
\\
d_{l_i}(N)
&:=
(-2N)^{l_i-1},
\end{align}
and where $\delta=\pm 1$ and $c=0,1/2$ in the Hermite and Laguerre cases respectively.

Our task now is to simplify the determinant appearing in (\ref{q-det form of lim F}) by using the contour integral 
(\ref{general form of the r derivatives}). This is achieved by the following lemma.
\begin{lemma}
If $q\in \mathbb{N}^m$ and $\delta=\pm 1$,
and we define
\begin{equation}
b_{\delta,q_i}(z):=
\begin{cases}
1&\delta=+1,\\
i \, z^{1-2q_i}& \delta=-1,
\end{cases}
\end{equation}
then 
\begin{multline}
\det_{\substack{1\le l_i \le 2q_i \\ 1\le i\le m\\1\le j\le 2|q|}}
\left[
\int_{\mathcal{C}}dz_{\dblqsum(i)+l_i} \,\,e^{-NS(z_{\dblqsum(i)+l_i},\mu_i)}
\,\afn(z_{\dblqsum(i)+l_i})\,z_{\dblqsum(i)+l_i}^{l_i-1}[z_{\dblqsum(i)+l_i}^{\delta}+c]^{j-1}
\right]
\\=
\prod_{i=1}^m \frac{1}{\Gamma(2q_i+1)}
\prod_{i=1}^m\prod_{l_i=\dblqsum(i)+1}^{\dblqsum(i+1)}
\int_{\mathcal{C}}dz_{l_i}\, e^{-N S(z_{l_i},\mu_i)}\, \afn(z_{l_i})\, b_{\delta,q_i}(z_{l_i})\, 
\prod_{i=1}^m \Delta_{2q_i}^2(z_{\dblqsum(i)+1},\dots,z_{\dblqsum(i+1)})
\\\times
\prod_{1\le j<k\le m}\,\prod_{l_k=\dblqsum(k)+1}^{\dblqsum(k+1)}\,\prod_{l_j=\dblqsum(j)+1}^{\dblqsum(j+1)}
(z_{l_k}^{\delta}-z_{l_j}^{\delta}).
\label{q-det equation}
\end{multline}
\label{q-det}
\end{lemma}
\begin{proof}
We start with the identity 
\begin{multline}
\det_{\substack{1\le l_i \le 2q_i \\ 1\le i\le m\\1\le j\le 2|q|}}
\left[\int dz_{\dblqsum(i)+l_i}\,\,
g_{l_i}(z_{\dblqsum(i)+l_i},\mu_i)[f(z_{\dblqsum(i)+l_i})]^{j-1}
\right]
\\
=\prod_{i=1}^m\prod_{l_i=1}^{2q_i} \int dz_{\dblqsum(i)+l_i}\,\,g_{l_i}(z_{\dblqsum(i)+l_i},\mu_i)
\Delta_{2|q|}\left( f(z_1),\dots,f(z_{2|q|})\right),
\label{det integral}
\end{multline}
which is valid for arbitrary integrable functions $f(z)$ and $g_{l_i}(z,\mu_i)$.
If we apply (\ref{det integral}) to the left hand side (LHS) of (\ref{q-det equation}) and use the elementary fact that
\begin{equation}
\Delta_n(z_1+c,z_2+c,\dots,z_n+c)=\Delta_n(z_1,z_2,\dots,z_n),
\end{equation}
we obtain
\begin{equation}
\text{LHS of (\ref{q-det equation})}
=
\prod_{i=1}^m\prod_{l_i=1}^{2q_i}
\int_{\mathcal{C}}dz_{\dblqsum(i)+l_i}\,z_{\dblqsum(i)+l_i}^{l_i-1}
\prod_{i=1}^m\prod_{l_i=\dblqsum(i)+1}^{\dblqsum(i+1)}
 e^{-N S(z_{l_i},\mu_i)}\, \afn(z_{l_i})\,
\mult\,\Delta_{2|q|}(z_{1}^{\delta},\dots,z_{2|q|}^{\delta}).
\label{q-det after determinantal integral lemma}
\end{equation}

To proceed further we first note the following two useful identities.
\begin{lemma} If $f(z_1,\dots,z_{2|q|})$ is a totally antisymmetric function of each set of 
variables $\{z_{\dblqsum(i)+1},\dots,z_{\dblqsum(i+1)}\}$, for $i=1,2,\dots,m$, then
\begin{multline}
\prod_{i=1}^m\,\prod_{l_i=1}^{2q_i}\int\,dz_{\dblqsum(i)+l_i}\,z_{\dblqsum(i)+l_i}^{l_i-1}\,\,
f(z_1,\dots,z_{2|q|})
\\=
\prod_{i=1}^m\frac{1}{\Gamma(2q_i+1)}
\prod_{i=1}^m\,\left(\prod_{l_i=\dblqsum(i)+1}^{\dblqsum(i+1)}\int\,dz_{l_i}\right)
\Delta_{2q_i}(z_{\dblqsum(i)+1},\dots,z_{\dblqsum(i+1)})
\,.\,f(z_1,\dots,z_{2|q|})
\label{hidden Vandermonde equation}
\end{multline}
\label{hidden Vandermonde}
\end{lemma}
\begin{proof}
By expanding the Vandermonde determinant and then rearranging the
order of integrations we see that
\begin{multline}
\prod_{i=1}^m\,\left(\prod_{l_i=\dblqsum(i)+1}^{\dblqsum(i+1)}\int\,dz_{l_i}\right)
\Delta_{2q_i}(z_{\dblqsum(i)+1},\dots,z_{\dblqsum(i+1)})\,.\,f(\dots,z_{\dblqsum(i)+1},\dots,z_{\dblqsum(i)+2q_i},\dots)
\\=
\prod_{i=1}^m\,
\sum_{\sigma_i\in S_{2q_i}}\,\prod_{l_i=1}^{2q_i}
\int\, dz_{\dblqsum(i)+\sigma_i(l_i)}\, z_{\dblqsum(i)+\sigma_i(l_i)}^{l_i-1}\,(-1)^{\sigma_i}\, .\,
f(\dots,z_{\dblqsum(i)+1},\dots,z_{\dblqsum(i)+2q_i},\dots),
\label{hidden Vandermonde proof 1}
\end{multline}
and the antisymmetry of $f$ then implies that the right hand side of (\ref{hidden Vandermonde proof 1}) equals
\begin{multline}
\prod_{i=1}^m\,
\sum_{\sigma_i\in S_{2q_i}}\,\prod_{l_i=1}^{2q_i}
\int\, dz_{\dblqsum(i)+\sigma_i(l_i)}\, z_{\dblqsum(i)+\sigma_i(l_i)}^{l_i-1}\, \mult\,
f(\dots,z_{\dblqsum(i)+\sigma_i(1)},\dots,z_{\dblqsum(i)+\sigma_i(2q_i)},\dots)
\\=
\prod_{i=1}^m
\sum_{\sigma_i\in S_{2q_i}}\,\prod_{l_i=1}^{2q_i}
\int\, dz_{\dblqsum(i)+l_i}\, z_{\dblqsum(i)+l_i}^{l_i-1}\, .\,f(\dots,z_{\dblqsum(i)+1},\dots,z_{\dblqsum(i)+2q_i},\dots),
\end{multline}
where the last equality follows by simply relabelling integration variables. The stated result is now immediate.
\end{proof}

\begin{lemma}
With $q$, $\delta$ and $b_{\delta,q_i}(z)$ as defined in Lemma \ref{q-det} we have
\begin{equation}
\Delta_{2q_i}
\left(
z_1^{\delta},\dots,z_{2q_i}^{\delta}
\right)
\Delta_{2q_i}
\left(
z_1,\dots,z_{2q_i}
\right)
=\prod_{l=1}^{2q_i}b_{\delta,q_i}(z_l)\,
\mult\,\Delta_{2q_i}^2(z_1,\dots,z_{2q_i}).
\label{Vandermonde product equation}
\end{equation}
\label{Vandermonde product}
\end{lemma}
\begin{proof}
When $\delta=1$ there is nothing to prove, so take $\delta=-1$. Then
\begin{align}
\Delta_{2q_i}
\left(
z_1^{\delta},\dots,z_{2q_i}^{\delta}
\right)
\Delta_{2q_i}
\left(
z_1,\dots,z_{2q_i}
\right)
&=
\prod_{1\le j<k\le 2q_i} \frac{(z_j-z_k)(z_k-z_j)}{z_j z_k},
\\
&=(-1)^{2q_i^2+q_i}\prod_{1\le j<k\le 2q_i}\frac{1}{z_j z_k}\,\mult\,\Delta_{2q_i}^2(z_1,\dots,z_{2q_i}),
\\
&=
\prod_{l=1}^{2q_i}i\, z_l^{1-2q_i}\,\,\mult\,\Delta_{2q_i}^2(z_1,\dots,z_{2q_i}).
\end{align}
\end{proof}
Armed with Lemmas \ref{hidden Vandermonde} and \ref{Vandermonde product} the proof of Lemma \ref{q-det} follows at once.
Applying Lemma \ref{hidden Vandermonde} to (\ref{q-det after determinantal integral lemma}) results in 
\begin{align}
\text{LHS of (\ref{q-det equation}})&=
\prod_{i=1}^m\frac{1}{\Gamma(2q_i+1)}
\prod_{i=1}^m
\left(
\prod_{l_i=\dblqsum(i)+1}^{\dblqsum(i+1)}
\int_{\mathcal{C}}
dz_{l_i}
\right)
\Delta_{2q_i}(z_{\dblqsum(i)+1},\dots,z_{\dblqsum(i+1)})
\nonumber\\
&\quad\times
\prod_{i=1}^m\,\prod_{l_i=\dblqsum(i)+1}^{\dblqsum(i+1)}
e^{-N S(z_{l_i},\mu_i)}\afn(z_{l_i})\,.\,\Delta_{2|q|}(z_1^{\delta},\dots,z_{2|q|}^{\delta})
\\
&=
\prod_{i=1}^m\frac{1}{\Gamma(2q_i+1)}\,
\prod_{i=1}^m\,\prod_{l_i=\dblqsum(i)+1}^{\dblqsum(i+1)}
\int_{\mathcal{C}}
dz_{l_i}
e^{-N S(z_{l_i},\mu_i)}\afn(z_{l_i})
\nonumber\\
&\quad\times
\prod_{i=1}^m
\Delta_{2q_i}(z_{\dblqsum(i)+1}^{\delta},\dots,z_{\dblqsum(i+1)}^{\delta})
\Delta_{2q_i}(z_{\dblqsum(i)+1},\dots,z_{\dblqsum(i+1)})
\nonumber
\\
&\quad\times
\prod_{1\le j<k\le m}\,
\prod_{l_k=\dblqsum(k)+1}^{\dblqsum(k+1)}
\,
\prod_{l_j=\dblqsum(j)+1}^{\dblqsum(j+1)}
(z_{l_k}^{\delta}-z_{l_j}^{\delta}).
\label{q-det after hidden Vandermonde and a rearrangement}
\end{align}
Applying Lemma \ref{Vandermonde product} to the right hand side of 
(\ref{q-det after hidden Vandermonde and a rearrangement}) produces the stated result.
\end{proof}

Now we substitute (\ref{general form of the r derivatives}) into (\ref{q-det form of lim F}),
factor out the constants $c_j(N)$ and $d_{l_i}(N)$ from the determinant, and apply Lemma \ref{q-det} to finally
 obtain
\begin{prop}
\label{H duality proposition}
\begin{equation}
{\mathcal{H}}_{N,m,q}(\mu)
=\prefac\,
\prod_{i=1}^m\frac{1}{G(2q_i+1)\Gamma(2q_i+1)}
\prod_{i=1}^m\zeta_N^{2q_i}(\mu_i)\,
\prod_{1\le j<k\le m}(\mu_k-\mu_j)^{-4 q_j q_k}\, \mult\,\I,
\label{H duality formula}
\end{equation}
where
\begin{equation}
\begin{split}
I_{N,m,q}(\mu)
&
:=\prod_{i=1}^m\prod_{l_i=\dblqsum(i)+1}^{\dblqsum(i+1)}\int_{\mathcal{C}}dz_{l_i}\,
e^{-N S(z_{l_i},\mu_i)}\,\Delta_{2q_i}^2(z_{\dblqsum(i)+1},\dots,z_{\dblqsum(i+1)})
\\&
\phantom{:}\quad\times
\prod_{i=1}^m\prod_{l_i=\dblqsum(i)+1}^{\dblqsum(i+1)}
g_{q_i}(z_{l_i})
\,\,\prod_{1\le j<k\le m}
\,\,
\prod_{l_k=\dblqsum(k)+1}^{\dblqsum(k+1)}
\,\,
\prod_{l_j=\dblqsum(j)+1}^{\dblqsum(j+1)}
(z_{l_k}^{\delta}-z_{l_j}^{\delta}),
\label{I definition}
\end{split}
\end{equation}
the function $g_{q_i}(z)$ is 
\begin{align}
g_{q_i}(z):&= \afn(z)\,b_{\delta,q_i}(z)
\\
&=
\begin{cases}
1 & \text{Hermite},\\
\begin{displaystyle}i \frac{(z+2)^{\alpha}}{z^{2q_i}}\end{displaystyle} &\text{Laguerre,}\\
\end{cases}
\end{align}
and 
\begin{equation}
\prefac:=\prod_{j=1}^{\dblqsum(m+1)}c_j(N)\,\prod_{i=1}^m\prod_{l_i=1}^{2q_i}d_{l_i}(N)\,
\mult\,\frac{H_{N,N}}{H_{N+|q|,N}}.
\label{prefac definition}
\end{equation}
\end{prop}
Proposition \ref{H duality proposition} is an exact duality formula when $q\in\mathbb{N}^m$, 
expressing the $N$-fold integral $\h$ in 
terms of the $2|q|$-fold integral $I_{N,m,q}(\mu)$.
This allows us to compute the large $N$ asymptotics of $\h$
by computing the large $N$ asymptotics of $I_{N,m,q}(\mu)$, and the
latter can be obtained by using the saddle point method. 
This is the subject of Section \ref{Asymptotics}.

The prefactor $\prefac$ defined in (\ref{prefac definition}) can be expressed in terms of the Barnes' $G$-function
by using known results for the Selberg Integral,
see e.g. \cite{Mehta,ForresterBook}, and the asymptotics can then be obtained from the known asymptotics of Barnes' 
$G$-function\cite{FerreiraLopez01}.
We obtain:
\begin{align}
\prefac
&=
\begin{cases}
\begin{displaystyle}
2^{-|q|^2-3|q|/2+\sum_{i=1}^m2q_i^2}\pi^{-3|q|/2}\,N^{\sum_{i=1}^m2q_i^2+|q|^2/2+|q| N}
\frac{G(N+2)}{G(N+|q|+2)}
\end{displaystyle}
& \text{Hermite},\\
 \begin{displaystyle}
{\left( -1 \right) }^{|q| }\,2^{-2\,|q|  + \sum_{i = 1}^{m}2\,{{q_i}}^2}\,
  {\pi }^{-2\,|q| }
\,  N^{\left( \alpha  - |q|  \right) \,|q|  + \sum_{i = 1}^{m}2\,{{q_i}}^2}\,
\end{displaystyle}
\\
\begin{displaystyle}
\quad\times
\frac{G(N+2)}{G(N+1)}
\frac{G(N + \alpha+1 )}{G(N + \alpha  + |q| +1)}
\frac{G(N + 2|q| +1)}{G(N + |q|+2 )}
\end{displaystyle}
 & \text{Laguerre},
\end{cases}
\\
&=
\begin{cases}
\begin{displaystyle}
N^{\sum_{i=1}^m(2q_i^2-q_i)}
e^{|q| N}2^{-|q|^2}\,
\prod_{i=1}^m 2^{2q_i^2-2q_i}\pi^{-2\,q_i}\,
\left[
1+O\left(\frac{1}{N}\right)
\right]
\end{displaystyle}
& \text{Hermite},\\
\begin{displaystyle}
N^{\sum_{i=1}^m(2q_i^2-q_i)}
\prod_{i=1}^m
(-1)^{q_i}
\,
2^{2\,{{q_i}}^2-2\,q_i}
\pi^{-2\,q_i }
\left[ 1
+O\left(\frac{1}{N}\right)
\right]
\end{displaystyle}
& \text{Laguerre}.
\end{cases}
\end{align}
It will be useful in Section \ref{Asymptotics} for us to introduce the notation
\begin{equation}
h_{N,m,q}=N^{\sum_{i=1}^m(2q_i^2-q_i)} \, h_0\left[1+O\left(\frac{1}{N}\right)\right],
\label{general form of h asymptotics}
\end{equation}
where 
\begin{equation}
h_0
:=
\begin{cases}
\begin{displaystyle}
e^{|q| N}2^{-|q|^2}\,
\prod_{i=1}^m 2^{2q_i^2-2q_i}\pi^{-2\,q_i}
\end{displaystyle}
& \text{Hermite},\\
\begin{displaystyle}
\prod_{i=1}^m
(-1)^{q_i}
\,
2^{2\,{{q_i}}^2-2\,q_i}
\pi^{-2\,q_i }
\end{displaystyle}
& \text{Laguerre}.
\end{cases}
\end{equation}

\section{Asymptotics}
\label{Asymptotics}
Now we begin the task of computing the large $N$ asymptotics of the
integral $I_{N,m,q}(\mu)$ for fixed $\mu$ and $q$. Since the
only appearance that $N$ makes in (\ref{I definition}) is in the
exponent of $e^{-N \,S(z,\mu_i)}$, this problem is a natural candidate
for the saddle point method.

In both the Hermite and Laguerre cases, the function $S(z,\mu_i)$ has
two saddle points, $\sad$ and its complex conjugate
$\sadstar$. Explicitly 
\begin{equation}
\sad = 
\begin{cases}
\mu_i +i \sqrt{1-\mu_i^2}& \text{Hermite,}\\
\begin{displaystyle}
 -1 + i\sqrt{\frac{1}{\mu_i}-1}
\end{displaystyle}
&\text{Laguerre.}\\
\end{cases}
\end{equation}
It is worth noting that in both cases 
\begin{equation}
\im\{\sad\}= \frac{\pi}{2}\,\rho(\mu_i),
\end{equation}
where $\rho(\mu_i)$ is as defined in (\ref{Hermite weight}) and
(\ref{Laguerre weight}), for the
Hermite and Laguerre cases respectively.
Both saddle points are of equal importance,
since
\begin{equation}
\re\{S(\sad,\mu_i)\}=\re\{S(\sadstar,\mu_i)\},
\end{equation}
and we deform the contour $\mathcal{C}$ through
both of them.

Let us denote the subset of the contour in
neighborhoods of $\sad$ and $\sadstar$ by $\nbhd$ and $\nbhdstar$ respectively, and the complement of
the union of these two neighborhoods in $\mathcal{C}$ by $\mathcal{C}_s$, so that 
\begin{equation}
\mathcal{C}=\mathcal{C}_s\cup\Omega_{+,i}\cup \Omega_{-,i}.
\end{equation}
By deforming $\mathcal{C}$ appropriately,
the dominant contribution of each integral comes from $\nbhd$ and $\nbhdstar$, and the standard arguments 
of the saddle point method lead to 
\begin{multline}
\I = \prod_{i=1}^m\,\prod_{l_i=\dblqsum(i)+1}^{\dblqsum(i+1)}
\left( 
\int_{\nbhd}dz_{l_i}+\int_{\nbhdstar}dz_{l_i}
\right)
e^{-N\,S(z_{l_i},\mu_i)}\mult\Delta_{2q_i}^2(z_{\dblqsum(i)+1},\dots,z_{\dblqsum(i+1)})
\\\times
\prod_{i=1}^m
\,
\prod_{l_i=\dblqsum(i)+1}^{\dblqsum(i+1)}g_{q_i}(z_{l_i})
\,
\prod_{1\le j<k\le m}
\,
\prod_{l_k=\dblqsum(k)+1}^{\dblqsum(k+1)}
\,
\prod_{l_j=\dblqsum(j)+1}^{\dblqsum(j+1)}
(z_{l_k}^{\delta}-z_{l_j}^{\delta})
+
\prod_{i=1}^m e^{-2q_i\,N \,\re\{S_i\}} \mult\,O(e^{-\varepsilon N}),
\label{I after throw away lemma}
\end{multline}
for suitably small $\varepsilon>0$, where we have defined
\begin{align}
S_i
&:=
S(\sad,\mu_i)=S(\sadstar,\mu_i)^*
.
\end{align}

We would now like to expand out the $2|q|$-fold composition of the sum of the two integrals appearing in 
(\ref{I after throw away lemma}). 
To achieve this, we first note that the integrand in (\ref{I after throw away lemma}) is totally 
symmetric in each set of variables
${\{z_{\dblqsum(i)+1},\dots,z_{\dblqsum(i+1)}\}}$, for $i=1,2,\dots,m$. With this in mind we can then apply 
the following elementary result.
\begin{lemma}
\label{binomial integral expansion}
If $f(z_1,\dots,z_{\dblqsum(i)+1},\dots,z_{\dblqsum(i+1)},\dots,z_{\dblqsum(m+1)})$ 
is a totally symmetric function of the
variables $\{z_{\dblqsum(i)+1},\dots,z_{\dblqsum(i+1)}\}$, 
then
\begin{multline}
\prod_{l_i=\dblqsum(i)+1}^{\dblqsum(i+1)}
\left(\int_{\nbhd}dz_{l_i}+\int_{\nbhdstar}dz_{l_i}
\right)
\,f(z_1,\dots,z_{\dblqsum(i)+1},\dots,z_{\dblqsum(i+1)},\dots,z_{\dblqsum(m+1)})
\\
=
\sum_{n_i=0}^{2q_i}\binom{2q_i}{n_i}
\prod_{l_i=\dblqsum(i)+1}^{\dblqsum(i)+n_i}
\int_{\nbhd}dz_{l_i} 
\prod_{l_i=\dblqsum(i)+n_i+1}^{\dblqsum(i+1)}
\int_{\nbhdstar}dz_{l_i}
\,\,
f(z_1,\dots,z_{\dblqsum(i)+1},\dots,z_{\dblqsum(i+1)},\dots,z_{\dblqsum(m+1)}).
\\
\end{multline}
\end{lemma}
\begin{proof}
All terms in the expansion with $n_i$ integrals over $\nbhd$ can be seen to be equal by swapping the order of the 
integrations, permuting the arguments of $f$ in ${\{z_{\dblqsum(i)+1},\dots,z_{\dblqsum(i+1)}\}}$,  and then 
relabelling the integration variables appropriately.
\end{proof} 
Applying  Lemma \ref{binomial integral expansion} to (\ref{I after throw away lemma}) results in
\begin{multline}
\I =
\prod_{i=1}^m \, 
\sum_{n_i=0}^{2q_i}\binom{2q_i}{n_i}
\,\,\prod_{l_i=\dblqsum(i)+1}^{\dblqsum(i)+n_i}
\int_{\nbhd}dz_{l_i} 
\prod_{l_i=\dblqsum(i)+n_i+1}^{\dblqsum(i+1)}
\int_{\nbhdstar}dz_{l_i}\,               
\\
\prod_{i=1}^m\,
\left(
\prod_{l_i=\dblqsum(i)+1}^{\dblqsum(i+1)} \,e^{-N\,S(z_{l_i},\mu_i)}\, g_{q_i}(z_{l_i})
\right)
\Delta_{2q_i}^2(z_{\dblqsum(i)+1},\dots,z_{\dblqsum(i+1)})
\mult
\prod_{1\le j<k\le m}
\prod_{l_k=\dblqsum(k)+1}^{\dblqsum(k+1)}
\prod_{l_j=\dblqsum(j)+1}^{\dblqsum(j+1)}
(z_{l_k}^{\delta}-z_{l_j}^{\delta})
\\+
\prod_{i=1}^m e^{-2q_i\,N \,\re\{S_i\}} \,\mult\,O(e^{-\varepsilon N}).
\label{I after binomial integral expansion}
\end{multline}

Now let us parameterize the integration variables in (\ref{I after binomial integral expansion}) so that the paths 
$\nbhdboth$ become line segments, (of length $2\eta$ say), centered at
$\sad$ and $\sadstar$ respectively and lying along the direction of steepest descent 
\begin{equation}
z_{l_i}=
\begin{cases}
\sad     +  e^{i\theta_i}\,t_{l_i}, & \dblqsum(i)+1\le l_i\le \dblqsum(i)+n_i, \\
\sadstar + e^{-i\theta_i}\,t_{l_i}, & \dblqsum(i)+n_i+1\le l_i\le \dblqsum(i)+2q_i.  \\
\end{cases}
\label{change of variables}
\end{equation}
The angles $\theta_i$ are chosen in the usual way so that with
\begin{equation}
a_i:=\frac{S''(\sad,\mu_i)}{2}e^{2\,i\,\theta_i}=\frac{S''(\sadstar,\mu_i)}{2}e^{-2\,i\,\theta_i}
\end{equation}
we have $a_i\in(0,\infty)$, and so
\begin{equation}
a_i \,t_{l_i}^2
=
\frac{S''(\sad,\mu_i)}{2}(z_{l_i}-\sad)^2=\frac{S''(\sadstar,\mu_i)}{2}(z_{l_i}-\sadstar)^2,
\end{equation}
with $t_{l_i}\in[-\eta,\eta]$. With this convention the integrals through
$\sad$ are oriented in the negative direction, and we will compensate
for this by introducing the explicit factor $(-1)^{n_i}$. Explicitly,
since
\begin{equation}
\frac{S''(\sad,\mu_i)}{2}=
\begin{cases}
\begin{displaystyle}
\frac{\pi}{2}
\,\rho(\mu_i)\,e^{i(\Arcsin(\mu_i) + \pi)}
\end{displaystyle}
& \text{Hermite,}\\
\begin{displaystyle}
\pi\,\mu_i^2\,
\rho(\mu_i)\,e^{-i\pi/2}
\end{displaystyle}
& \text{Laguerre,}\\
\end{cases}
\end{equation}
we obtain
\begin{equation}
\theta_i
=
\begin{cases}
\begin{displaystyle}
\frac{\pi-\Arcsin(\mu_i)}{2}
\end{displaystyle}
& \text{Hermite},\\
\begin{displaystyle}
\frac{\pi}{4}
\end{displaystyle}
& \text{Laguerre},\\
\end{cases}
\end{equation}
\begin{equation}
a_i
=
\begin{cases}
\begin{displaystyle}
\frac{\pi}{2}\,\rho(\mu_i) 
\end{displaystyle}
& \text{Hermite},\\
\pi\,\mu_i^2\,\rho(\mu_i) & \text{Laguerre}.\\
\end{cases}
\end{equation}
Note that $a_i>0$ when $\mu_i\in \Int(\text{supp}\,\rho)$.

In making the change of variables (\ref{change of variables}) in (\ref{I after binomial integral expansion}),
it is convenient to introduce the the following definitions:
\begin{align}
\phi_{N,i}
&:=
4q_i\,\theta_i+2N\,\im\{S_i\},
\\
\varphi_i(t_{l_i})
&:=
\sum_{k=1}^{\infty}\frac{2}{S''(z_{+,i},\mu_i)}\,
\frac{S^{(k+2)}(z_{+,i},\mu_i)}{(k+2)!}
 (z_{l_i}-z_{+,i})^k\,
\Big\vert_{z_{l_i} \,=\, z_{+,i}+e^{i\theta_i t_{l_i}}},
\\
\Phi_{{ n}}(t)
&:=
\prod_{i=1}^m
\prod_{l_i=\dblqsum(i)+1}^{\dblqsum(i)+n_i}\,e^{-N a_i\, t_{l_i}^2\varphi_i(t_{l_i})}
\prod_{l_i=\dblqsum(i)+n_i+1}^{\dblqsum(i+1)}\,e^{-N a_i\, t_{l_i}^2{\varphi}^*_i(t_{l_i})},
\\
G_{{ n}}(t)
&:=
\prod_{i=1}^m
\prod_{l_i=\dblqsum(i)+1}^{\dblqsum(i+1)}g_{q_i}(z_{l_i})\Big\vert_{\star},
\label{G definition}
\\
D_{{ n}}(t)
&:=
\prod_{i=1}^m
\prod_{k_i=\dblqsum(i)+n_i+1}^{\dblqsum(i+1)}
\prod_{j_i=\dblqsum(i)+1}^{\dblqsum(i)+n_i}
(z_{k_i}-z_{j_i})^2\Big\vert_{\star},
\label{D definition}
\\
H_{{ n}}(t)
&:=
\prod_{1\le j<k\le m}
\prod_{l_k=\dblqsum(k)+1}^{\dblqsum(k+1)}
\prod_{l_j=\dblqsum(j)+1}^{\dblqsum(j+1)}
(z_{l_k}^{\delta}-z_{l_j}^{\delta})\Big\vert_{\star},
\label{H definition}
\\
F_{{ n}}(t)
&:=G_{{n}}(t)D_{{n}}(t)H_{{n}}(t),
\end{align}
where $\star$ on the right hand sides of (\ref{G definition}), 
(\ref{D definition}) and (\ref{H definition}) refers to the change of
variables (\ref{change of variables}), and $n:=(n_1,\dots,n_m)$.
We also define the (unnormalized) integral operator
\begin{equation}
\mathbb{E}^{(i)}_{r,p}
:= 
\prod_{l_i=\dblqsum(i)+r}^{\dblqsum(i)+p}\int_{-\eta}^{\eta}dt_{l_i}\,e^{-N\,a_i\,t_{l_i}^2}
\,\,
\Delta_{p-r+1}^2(t_{\dblqsum(i)+r},\dots,t_{\dblqsum(i)+p}).
\label{integral operator}
\end{equation}
Armed with these definitions, we see that since 
\begin{multline}
\Delta_{n_i}^2(z_{\dblqsum(i)+1},\dots,z_{\dblqsum(i)+n_i})
\,
\Delta_{2q_i-n_i}^2(z_{\dblqsum(i)+n_i+1},\dots,z_{\dblqsum(i)+2q_i})
\,
\prod_{l_i=\dblqsum(i)+1}^{\dblqsum(i+1)}dz_{l_i}\,
\\=
e^{i 4q_i(q_i-n_i)\theta_i}
\Delta_{n_i}^2(t_{\dblqsum(i)+1},\dots,t_{\dblqsum(i)+n_i})
\,\,
\Delta_{2q_i-n_i}^2(t_{\dblqsum(i)+n_i+1},\dots,t_{\dblqsum(i)+2q_i})
\prod_{l_i=\dblqsum(i)+1}^{\dblqsum(i+1)}dt_{l_i},
\end{multline}
and 
\begin{equation}
\prod_{i=1}^m\prod_{l_i=\dblqsum(i)+1}^{\dblqsum(i+1)}\,
e^{-N  S(z_{l_i},\mu_i)}
=\prod_{i=1}^m e^{-2q_i N \,\re\{S_i\}}
\,
\prod_{i=1}^m
\,
e^{2i(q_i-n_i)N\,\im\{S_i\}}
\,
\prod_{i=1}^m\prod_{l_i=\dblqsum(i)+1}^{\dblqsum(i+1)}
\,
e^{-N a_i t_{l_i}^2}
\,\,
\Phi_{{n}}(t),
\end{equation}
we arrive at the following more compact expression for $\I$
\begin{multline}
\I =\prod_{i=1}^me^{-2q_i\,N\,\re\{S_i\}}
\\
\times
\left[
\prod_{i=1}^m \, \sum_{n_i=0}^{2q_i}
\binom{2q_i}{n_i}
\,(-1)^{n_i}\,e^{i(q_i-n_i)\phi_{N,i}}\,
\mathbb{E}^{(i)}_{1,n_i}
\mathbb{E}^{(i)}_{n_i+1,2q_i}
\Phi_n(t)\,F_{{n}}(t)
+O(e^{-\varepsilon N})
\right].
\label{compact expression for I}
\end{multline}
The factor $(-1)^{n_i}$ results from the fact that we traversed the line through $\sad$ in the 
negative direction, whereas the line through $\sadstar$ is traversed in the positive direction.

To obtain an asymptotic expansion of $\I$ from 
(\ref{compact expression for I}) we proceed in direct analogy with the 
one-dimensional saddle point method (see e.g. \cite{Wong}) and introduce
the following generalization of $\Phi_n(t)$,
\begin{equation}
\Phi_n(t,u)
:=
\prod_{i=1}^m
\prod_{l_i=\dblqsum(i)+1}^{\dblqsum(i)+n_i}\,e^{u_{l_i}\varphi_i(t_{l_i})}
\prod_{l_i=\dblqsum(i)+n_i+1}^{\dblqsum(i+1)}\,e^{u_{l_i}{\varphi}^*_i(t_{l_i})},
\label{generalised Phi}
\end{equation}
where $u\in\mathbb{R}^{2|q|}$. For convenience we also
introduce the function
\begin{equation}
Q_n(t,u):=\Phi_n(t,u)\,F_n(t),
\end{equation}
so that with 
\begin{equation}
u_{l_i}=-N a_i t_{l_i}^2 \quad \text{for each} \quad 1\le l_i\le 2|q|,
\label{Wong's dirty trick}
\end{equation}
we have
\begin{equation}
Q_n(t,\{-N a_i t_{l_i}^2\})=\Phi_n(t)\,F_n(t).
\end{equation}

Let us suppose for the present that $u$ is some arbitrary fixed parameter
independent of $t$, and
consider the $k$th degree Taylor polynomial of $Q_n(t,u)$ as a
function of $t\in\mathbb{R}^{2\vert q\vert}$
\begin{equation}
Q_n(t,u)=\sum_{0\le |\alpha|\le k}
\frac{1}{\alpha!}\,
\left[
\frac{\partial^{\alpha_1}}{\partial t_1^{\alpha_1}}
\dots 
\frac{\partial^{\alpha_{2|q|}}}{\partial t_{2|q|}^{\alpha_{2|q|}}}
Q_n(t,u)\Big\vert_{t=0}
\right]
\,t^{\alpha}
\,+\,O(t^{\alpha})\Big\vert_{|\alpha|=k+1},
\label{Taylor polynomial for Q(t,u)}
\end{equation}
where $\alpha\in \mathbb{Z}_{\ge 0}^{2|q|}$ and we use the standard notations
$\alpha!=\alpha_1!\dots\alpha_{2|q|}!$
and 
$t^{\alpha}= t_1^{\alpha_1}\dots t_{2|q|}^{\alpha_{2|q|}}$
and $|\alpha|=\alpha_1+\dots+\alpha_{2|q|}$. 

If in (\ref{Taylor polynomial for Q(t,u)}) we now chose $u$ according to
(\ref{Wong's dirty trick})
we obtain
\begin{equation}
\begin{split}
Q_n(t,\{-N a_i t_{l_i}^2\})
&=
\sum_{0\le |\alpha|\le k}
\frac{1}{\alpha!}\,
\left[
\frac{\partial^{\alpha_1}}{\partial t_1^{\alpha_1}}
\dots 
\frac{\partial^{\alpha_{2|q|}}}{\partial t_{2|q|}^{\alpha_{2|q|}}}
\Phi_n(t,u)\,F_n(t)\Big\vert_{t=0}
\right]\Bigg\vert_{\{u_{l_i}=-N a_i t_{l_i}^2\}}
\,t^{\alpha}
\\
&\quad +
\,O(t^{\alpha})\Big\vert_{|\alpha|=k+1},
\end{split}
\label{Wong Taylor polynomial}
\end{equation}
where we emphasize that the partial derivatives with respect to $t$ on
the right hand side of (\ref{Wong Taylor polynomial}) are performed with $u$ fixed {\em before} making
the substitution (\ref{Wong's dirty trick}). The effect of
constructing the Taylor series in this way is that 
when (\ref{Wong Taylor polynomial}) is substituted into 
(\ref{compact expression for I}) and the integrations are performed, each term
corresponding to a given value of $|\alpha|$ in (\ref{Wong Taylor polynomial}) will
have the same $N$ dependence. 
To see this we need to consider 
the asymptotics of the integral operator (\ref{integral operator})
acting on a general monomial,
which can be deduced simply by scaling $N$ out of the
integral. We thus deduce that
\begin{equation}
\prod_{i=1}^m
\mathbb{E}_{1,n_i}^{(i)}\,\mathbb{E}_{n_i+1,2q_i}^{(i)}
\,\,t^{\alpha}
=
O\left(N^{-\sum_{i=1}^m q_i^2 \,-\,\sum_{i=1}^m(n_i-q_i)^2\,-\,|\alpha|/2}\right).
\label{asymptotic growth of Hermite Selberg moments}
\end{equation}
As a direct consequence of 
(\ref{asymptotic growth of Hermite Selberg moments}), we see that 
 with $u_{l_i}=- a_i N t_{l_i}^2$ for each $1\le l_i\le 2|q|$, we have 
\begin{equation}
\prod_{i=1}^m
\mathbb{E}_{1,n_i}^{(i)}\,\mathbb{E}_{n_i+1,2q_i}^{(i)}
\,\,t^{\alpha} \, u^{\alpha'}
=
O\left(N^{-\sum_{i=1}^m q_i^2 \,-\,\sum_{i=1}^m(n_i-q_i)^2\,-\,|\alpha|/2}\right)
\label{asymptotic growth equivalence}
\end{equation}
for any $\alpha_1',\dots,\alpha_{2|q|}'\in \mathbb{N}$; i.e. the left hand sides of
(\ref{asymptotic growth of Hermite Selberg moments}) and
(\ref{asymptotic growth equivalence}) have precisely the same
asymptotic dependence on $N$. 
From (\ref{generalised Phi}) its clear that when computing the partial derivatives of 
$\Phi_n(t,u)\,F_n(t)$ for a given $|\alpha|$ various powers of $u$
will appear, which after the substitution (\ref{Wong's dirty trick})
will mean we need to  calculate 
quantities of the form (\ref{asymptotic growth equivalence}).
However (\ref{asymptotic growth of Hermite Selberg moments}) 
and (\ref{asymptotic growth equivalence}) tell us that 
all such quantities have the same $N$ dependence. 
This would not be the case if we had just naively
constructed the Taylor polynomial of $\Phi_n(t)F_n(t)$. We remark
that since the symmetry of the integral operator (\ref{integral operator}) implies $t^{\alpha}$ is annihilated
whenever $|\alpha|$ is odd, only integer powers
of $1/N$ actually appear in the asymptotic expansion of $\I$, despite the
appearance of $|\alpha|/2$ in the exponent in (\ref{asymptotic growth equivalence}).

As a result of the expression (\ref{asymptotic growth equivalence}) we
can see that the dominant term in the asymptotic expansion of 
(\ref{compact expression for I}) occurs when both $n_i=q_i$ for all
$i=1,\dots,m$, and $|\alpha|=0$ in the Taylor expansion 
(\ref{Wong Taylor polynomial}). In general, the coefficient of the term
which is of order $1/N^k$ relative to the
leading term is composed of all terms for which
\begin{equation}
\sum_{i=1}^m(n_i-q_i)^2\,+\,\frac{|\alpha|}{2}=k.
\end{equation}

If we are interested only in retaining the leading term
the preceding arguments imply that
\begin{equation}
\I=\prod_{i=1}^m\,
\frac{e^{-2q_i\,N\,\re\{S_i\}}}{N^{q_i^2}}\,
\mult 
I_0
\left[
1+O\left(\frac{1}{N}\right)
\right],
\label{general form of I asymptotics}
\end{equation}
where the coefficient $I_0$ depends on $\mu$ and $q$
but is independent of $N$.
To obtain the explicit form of $I_0$ we first note that, since
$\Phi_n(0)=1$, we have 
\begin{equation}
\mathbb{E}^{(i)}_{r+1,r+p} \, \Phi_n(0) \,F_n(0)=\frac{1}{N^{p^2/2}}F_n(0)\, Z_p(a_i)\,+\,O(e^{-\eta^2 a_i N}),
\label{expectation normalisation}
\end{equation}
where 
\begin{align}
\label{Hermite Selberg moments}
Z_p(a_i)
&:=\int_{\mathbb{R}^p}d^p x \, \Delta_p^2(x)\, \prod_{l=1}^pe^{-a_i\, x_l^2},
\\
&\phantom{:}=
\left(
\frac{\pi}{2^{p-1}}
\right)^{p/2}
\frac{G(p+2)}{a_i^{p^2/2}}.
\label{explicit partition function}
\end{align}
The quantity $Z_p(a_i)$ is the normalization of the joint eigenvalue
pdf of the GUE and 
can be expressed in terms of the Selberg integral; see \cite{Mehta,ForresterBook}.
If we substitute (\ref{Wong Taylor polynomial}) into 
(\ref{compact expression for I}) and take the leading term, which corresponds to 
$\sum_{i=1}^m(n_i-q_i)^2 +|\alpha|/2=0$, then applying (\ref{expectation normalisation})
we find that
\begin{equation}
I_0=\prod_{i=1}^m\,\binom{2q_i}{q_i}\,(-1)^{q_i}\,Z_{q_i}^2(a_i)\,\mult\,F_q(0).
\label{I_0 definition}
\end{equation}
The explicit form of $F_q(0)$ can be obtained from the following results
\begin{align}
D_{{q}}(0)
&
=(-1)^{|q|}\,\pi^{\sum_{i=1}^m2q_i^2}\,\prod_{i=1}^m\rho^{2q_i^2}(\mu_i),
\label{D_q(0)}
\\
G_{q}(0)
&=
\begin{cases}
1,& \text{Hermite},\\
\begin{displaystyle}
\prod_{i=1}^m(-1)^{q_i}\,\mu_i^{2q_i^2}\,\mu_i^{-\alpha\,q_i}
\end{displaystyle}
,& \text{Laguerre},
\end{cases}
\label{G_q(0)}
\\
H_{{q}}(0)
&=\prod_{1\le j<k\le m}(\mu_k-\mu_j)^{2\,q_j\,q_k}\times
\begin{cases}
\begin{displaystyle}
2^{|q|^{2-\sum_{i=1}^mq_i^2}}
\end{displaystyle}
,& \text{Hermite},\\
1
,& \text{Laguerre}
.\\
\end{cases}
\label{H_q(0)}
\end{align}

Now let us put together what we have learned about the asymptotic
behavior of $\I$ to describe 
the asymptotic behavior of $\h$ when $q\in \mathbb{N}^m$. To this end, we
substitute (\ref{D_q(0)}) into (\ref{I_0 definition}), 
then (\ref{I_0 definition}) into 
(\ref{general form of I asymptotics}), and finally substitute
(\ref{general form of I asymptotics})  and (\ref{general form of h asymptotics}) into
(\ref{H duality formula}), to obtain
\begin{multline}
\h
=
N^{\sum_{i=1}^mq_i(q_i-1)}
\,
\prod_{i=1}^m\frac{G^2(q_i+1)}{G(2q_i+1)}2^{-q_i^2+q_i}\pi^{2q_i^2+q_i}\,\rho^{2q_i^2}(\mu_i)
\prod_{1\le j<k\le m}(\mu_k-\mu_j)^{-4\,q_j\,q_k}
\\\times
\prod_{i=1}^m e^{-2q_i\, N\,\re\{S_i\}}
\zeta_N^{2q_i}(\mu_i)\,a_i^{-q_i^2}\,\mult
G_{q}(0)H_{q}(0)\mult h_0
\left[1+O\left(\frac{1}{N}\right)\right].
\label{H with nonuniversal bits left}
\end{multline}

Using the explicit forms for $G_q(0)$ and $H_q(0)$, given by
(\ref{G_q(0)}) and  (\ref{H_q(0)}) respectively,
one can easily verify that for both the Hermite and Laguerre cases we
have the following identity
\begin{multline}
\prod_{i=1}^m e^{-2q_i\, N\,\re\{S_i\}}
\zeta_N^{2q_i}(\mu_i)\,a_i^{-q_i^2}\,\mult
  G_{q}(0)H_{q}(0)\mult h_0
\\
=\prod_{i=1}^m\omega_N^{-q_i}(\mu_i)
\,\rho^{-q_i^2}(\mu_i)\,2^{2q_i^2-2q_i}\,\pi^{-q_i^2-2q_i}\,
\prod_{1\le j<k\le m}(\mu_k-\mu_j)^{2q_j\,q_k}.
\label{universality after all}
\end{multline}

Inserting the identity (\ref{universality after all}) into 
(\ref{H with nonuniversal bits left}) we finally obtain 
\begin{equation}
\begin{split}
\h
&
=N^{\sum_{i=1}^m(q_i^2-q_i)}
\prod_{i=1}^m[\omega_N(\mu_i)]^{-q_i}
\prod_{i=1}^m 
\frac{G^2(q_i+1)}{G(2q_i+1)}(2\pi)^{q_i^2-q_i}
\\&\quad\times
\prod_{1\le j<k\le m}\vert \mu_k-\mu_j\vert^{-2q_j q_k}\prod_{l=1}^m[\rho(\mu_i)]^{q_i^2}
\left[1+O\left(\frac{1}{N}\right)\right],
\end{split}
\label{FF conjecture plus correction}
\end{equation}
which does indeed recover 
(\ref{FF conjecture equation}).

We emphasize that the derivation we have presented for
(\ref{FF conjecture plus correction}) is
entirely rigorous for $q\in\mathbb{N}^m$ for any $m\in\mathbb{N}$, thus verifying the legitimacy of the
log-gas procedure used in \cite{ForresterFrankel04} for the Hermite and Laguerre cases, for such $q$. 

\begin{acknowledgments}
I would like to thank Ofer Zeitouni and Greg Anderson for numerous helpful conversations in the early stages of this 
investigation, and Norm Frankel, Peter Forrester and Jon Keating for helpful
comments regarding an earlier draft. 
I also thank Dimitri Gangardt for encouraging me to pursue this work.
\end{acknowledgments}

\newpage

\end{document}